\documentclass{article}
\usepackage{spconf,amsmath,graphicx}
\usepackage{threeparttable}
\usepackage{float}
\usepackage{subfigure}
\usepackage{graphicx}
\usepackage{multirow}
\usepackage{booktabs}
\usepackage{tabularx}
\usepackage[hidelinks]{hyperref}
\usepackage{amssymb}
\usepackage{makecell}
\usepackage{epstopdf}
\title{MULTI-TASK SUB-BAND NETWORK FOR DEEP RESIDUAL ECHO SUPPRESSION}
\name{Jiayao Sun$^{1,2}$\thanks{This work was done during the internship at Li Auto. $*$: Corresponding author.}, Dawei Luo$^{2,*}$, Zhaoxia Li$^2$, Jindong Li$^2$, Yukai Ju$^{1}$, Yang Li$^2$}
\address{$^1$ Northwestern Polytechnical University, Xi'an, China\\
$^2$Li Auto, China
}
\begin{document}
\ninept
\maketitle
\begin{abstract}
\vspace{-2pt}
% Acoustic echo cancellation is a common problem in conferencing systems and communications. 
% 写我们的方法，不是竞赛的系统描述
This paper introduces the SWANT team’s entry to the ICASSP 2023 AEC Challenge. We submit a system that cascades a linear filter with a neural post-filter. Particularly, we adopt sub-band processing to handle full-band signals and shape the network with multi-task learning, where dual signal voice activity detection (DSVAD) and echo estimation are adopted as auxiliary tasks. 
Moreover, we particularly improve the time frequency convolution module (TFCM) to increase the receptive field using small convolution kernels.
% Meanwhile, this work improves time frequency convolution module (TFCM) for increasing the receptive field with small convolution kernels.
% Meanwhile, this work improve time frequency convolution module(TFCM) modeling capabilities between harmonics.
Finally, our system has ranked 4th in ICASSP 2023 AEC Challenge Non-personalized track.

\end{abstract}
\vspace{-2pt}
\vspace{-8pt}
\section{Introduction}
\label{sec:intro}
\vspace{-4pt}
% Recently, the processing of full-band voice has been a hot task, and the ICASSP 2022 AEC Challenge has promoted research on full-band AEC processing. Compared to 16K audio processing, 48K audio processing is more challenging, mainly for high-frequency processing. In [4], Zhang et al. processed only the wide-band (16Hz) signal, using the average gain control to synthesize the full-band signal. In [5], Zhao et al. processes the wideband signal and highband signal with wideband net and highband network, respectively. [6] Sun et al. processes the fullband signal directly.

% In this paper, we will introduce the system submitted to ICASSP 2023 Acoustic Echo Cancellation Challenge track1 (non-personalized), including gccphat delay estimation, NLMS linear filtering, and post-filtering with sub-band TFGCRN for residual echo suppression. 
Neural residual echo suppression~\cite{zhang2022multi} has achieved excellent performance in the AEC task. With the great success in previous challenges, the 4th AEC Challenge in ICASSP2023~\cite{cutler2023icassp} has particularly focused on more difficult \textit{full-band} signals.
To address this challenge with lower complexity, we propose a \textit{sub-band} time frequency domain gated convolutional recurrent neural network (S-TFGCRN) approach.  
% Compared with directly processing the full-band speech, pseudo quadrature mirror filter bank (PQMF) is used to process the waveform into sub-band speech, which can reduce computational complexity.
After linear filtering, the full-band signals are first processed to several sub-band signals using the pseudo quadrature mirror filter bank (PQMF) and the sub-band signals are fed separately into an encoder-decoder network to remove echo residuals. Finally, the sub-band signals are merged back into the full-band echo-removed signals. Importantly, our network is optimized under a multi-task learning framework, where dual signal VAD and echo estimation are augmented as auxiliary tasks. To better model the intrinsic relationships between harmonics, we introduce an updated time frequency convolution module (U-TFCM) in the encoder network.
%Additionally, the proposed updated TFCM (U-TFCM) is to better model the information between harmonics.
%Moreover, We employ a multi-task learning framework, including dual signal VAD and echo estimation modules as auxiliary tasks. Through an extensive ablation study, the UTFCM and the multi-task learning framework improved model performance.
% and these two modules can be removed during inference. 
% Finally, asymmetric loss is used to prevent near-end speech over-inhibition. 
The effectiveness of the contributions is validated through an ablation study. According to the official results of the challenge, our model ranks 4th in the non-personalized AEC track.

\vspace{-8pt}
%%%%%%%%%%%%% aec system图
\begin{figure}[h]
\vspace{-5pt}
\centering
\includegraphics[width=1\linewidth]{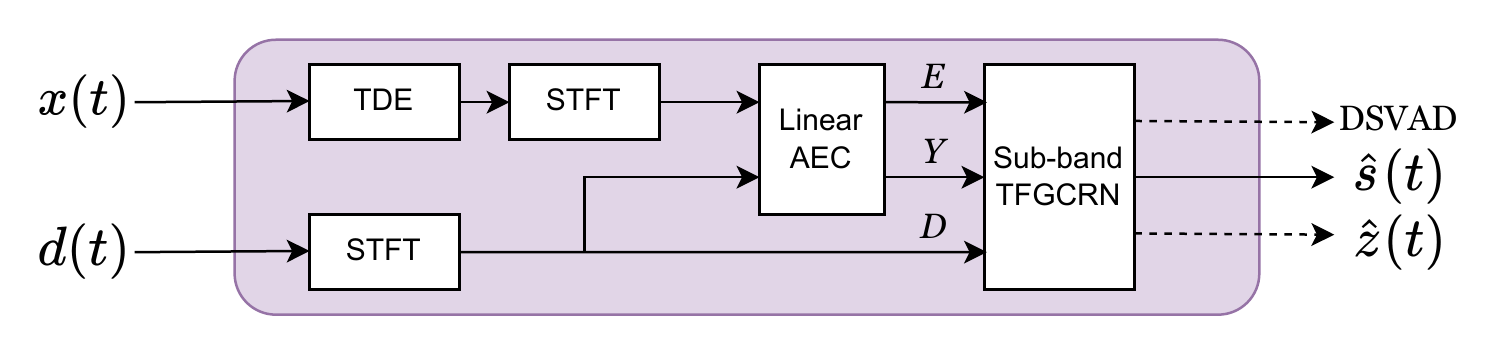}
\vspace{-20pt}
\caption{Architecture of our proposed AEC system.} 
\label{fig:aec}
\vspace{-20pt}
\end{figure}
%%%%%%%%%%%%%%%%%%%%%%%
% \vspace{-0.2cm}

\vspace{-8pt}

\section{PROPOSED SYSTEM}
\label{sec:pagestyle}
\vspace{-6pt}

\subsection{Problem formulation}
\vspace{-3pt}
The overall framework of our proposed AEC system is depicted in Fig.~\ref{fig:aec}, 
% For acoustic echo cancellation system, usually has two input signals, the near-end microphone signal $d(t)$ and the far-end microphone signal $x(t)$. 
where the near-end microphone signal $d(t)$ is a combination of several signals described as follows.
{
\setlength\abovedisplayskip{2.0pt}
\setlength\belowdisplayskip{2.0pt}
% \begin{equation}
\begin{align}
% \vspace{-0.1cm}
\footnotesize
    d(t) &= s(t) + z(t) + v(t) \\
    z(t) &= x(t) * h(t)
% \vspace{0.1cm}
\end{align}
}%
where $s(t)$, $x(t)$, $v(t)$ and $z(t)$ denote the near-end speech signal, far-end microphone signal, background noise, and echo signal, respectively, which is generated by the convolution of far-end signal and echo path $h(t)$. The t refers to the time sample index. This task is considered an audio separation task, and the goal is to separate the near-end speech signal $s(t)$ from the near-end microphone signal $d(t)$ and pass it to the far end. 
The echo $y$ is estimated by adaptive filter NLMS and the signal $e$ is the filter output.
$D$, $E$, $Y$, and $Z$ are frequency domain representations of $d$, $e$, $y$, and $z$, respectively. The time delay estimation (TDE) is implemented using Generalized Cross Correlation with PHAse Transform (GCC-PHAT). The dotted line in Fig.~\ref{fig:aec} indicates that the modules are used as auxiliary tasks and can be removed during model inference.

\begin{figure}[h]
\vspace{-10pt}
\centering
\includegraphics[width=1\linewidth]{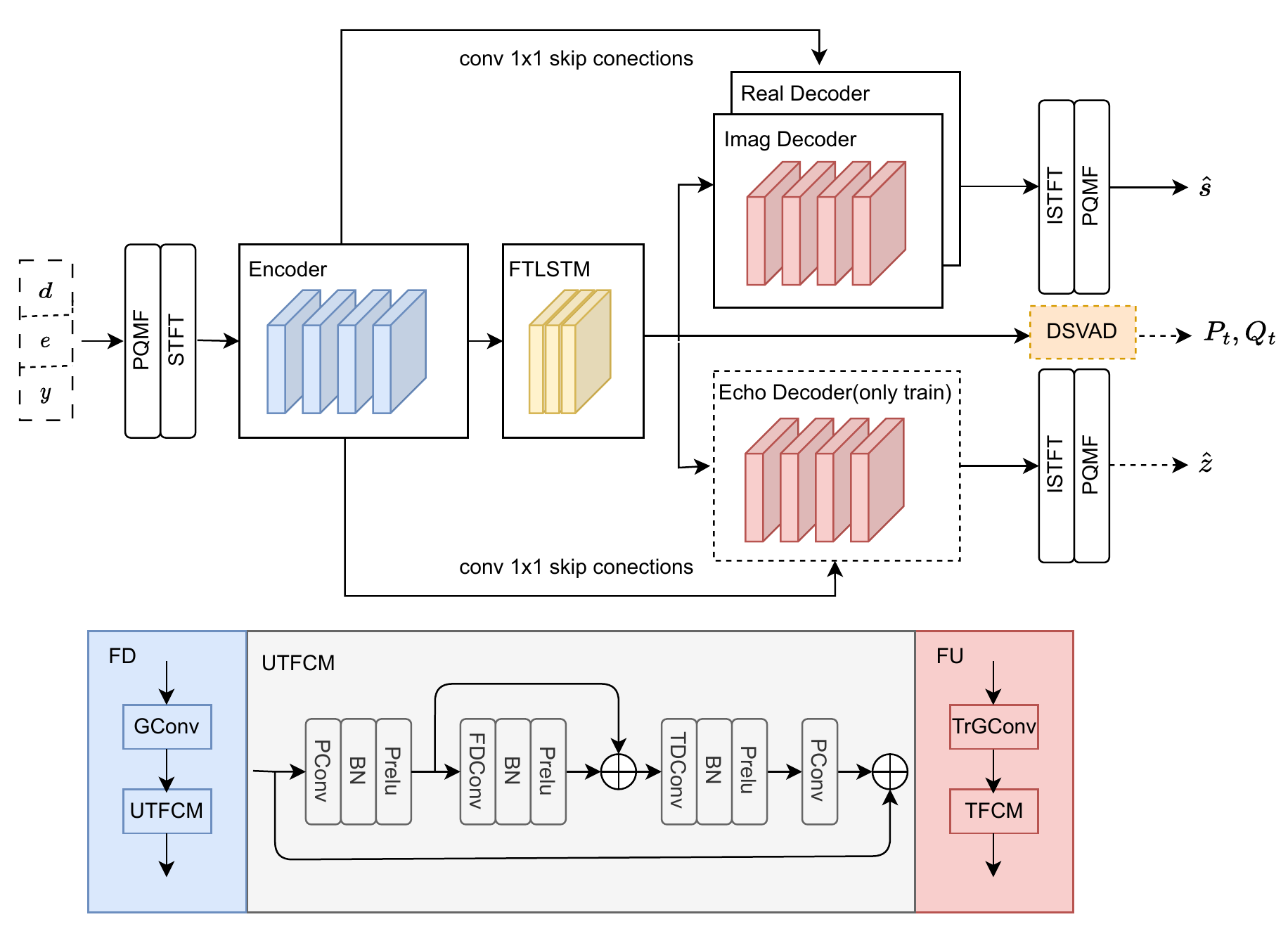}
\vspace{-20pt}
\caption{S-TFGCRN architecture.}
\vspace{-2pt}
%节省空间
\label{fig:all}
\vspace{-15pt}
\end{figure}
%%%%%%%%%%%%%%%%%%%%%%%
\vspace{-4pt}
\subsection{S-TFGCRN based post-filter} 
\vspace{-3pt}
We use the PQMF for signal analysis and synthesis. 
The sub-band analysis process comprises three steps:
% The process of sub-band analysis includes three parts: 
FIR analysis, downsampling, and STFT. 
Likewise, sub-band synthesis involves three stages 
% Sub-band synthesis also consists of three parts: 
ISFTT, upsampling, and FIR synthesis.

As illustrated in Fig.~\ref{fig:all}, the encoder part contains four frequency downsampling (FD) layers, and every FD layer contains GConv and U-TFCM modules. Different from the original TFCM~\cite{ju2023tea} that mainly focuses on the receptive field of the time dimension, the updated time frequency convolution module (U-TFCM) particularly increases the receptive field of the frequency dimension to learn time-frequency correlation. The skip connection uses a 1x1 convolution. FTLSTM is used as the bottleneck layer, which is proven to be effective in temporal modeling. The structures of GConv and FTLSTM follow those in~\cite{zhang2022multi}. The real/imag decoder part comprises four frequency upsampling (FU) layers, each containing a TrGConv module and a TFCM module. The output of the FU layer is concatenated with the output of the conv1x1. The echo decoder includes 4 FU layers, and its final TrGConv output layer has a channel dimension of 2. Estimating the echo target helps obtain a more accurate estimate of the near-end speech, as we treat the AEC task as an audio separation task.
% The structure of the echo decoder is the same as real/imag decoder, except that the channel dimension of the last layer of TrGConv output is 2. The echo decoder helps the network suppress echoes more effectively.
% Our previous work~\cite{zhang2022multi} only estimated the state of the near-end speech. In this work,  the state estimation of the far-end speech is increased, 
% so as to achieve distinguish between double-talk and single-talk scenarios. The DTVAD consists of two same VAD modules implemented in~\cite{zhang2022multi}. 
To distinguish between double-talk and single-talk scenarios, we do VAD processing for both near-end and far-end speech, resulting in dual signal VAD (DSVAD) that consists of two VAD modules. The structure of VAD follows that in~\cite{zhang2022multi}.
% As described in the previous work~\cite{zhang2022multi}, the same VAD structure is employed, but with an additional estimation of the probability of far-end speech in addition to the probability of near-end speech.
% the only difference is that we not only estimate the probability of near-end speech but also the probability of far-end speech.

\vspace{-10pt}
\subsection{Loss Function}
\vspace{-3pt}
Our loss function is based on the echo signal power weighted loss~\cite{zhang2022multi}, denoted as $\mathcal{L}_{\mathrm{echo-aware}}$. To ensure the accuracy of echo estimation for the echo decoder, $\mathcal{L}_{\mathrm{echo}}$ is defined as
{
\setlength\abovedisplayskip{2.0pt}
\setlength\belowdisplayskip{2.0pt}
\begin{equation}
\footnotesize
     \setlength{\arraycolsep}{0.3pt}
     \mathcal{L}_{\mathrm{echo}}=\sum_{f, t}|\widehat{Z}-Z|, \\
\end{equation}
}%
which is based on the mean absolute error (MAE) in ~\cite{braun2022task}. 
The binary cross entropy (BCE) is used as the DSVAD loss, formulated as
{
\setlength\abovedisplayskip{2.0pt}
\setlength\belowdisplayskip{2.0pt}
\begin{equation}
\footnotesize
% \left \{
     \setlength{\arraycolsep}{0.3pt}
     \begin{split}
     \mathcal{L}_{\mathrm{dtd-nearend}}&=\frac{1}{T} \underset{T}\sum\left(-\bar{P}_{t} \log \left(P_{t}\right)-\left(1-\bar{P}_{t}\right) \log \left(1-P_{t}\right)\right) \\
     \mathcal{L}_{\mathrm{dtd-farend}}&=\frac{1}{T} \underset{T}\sum\left(-\bar{Q}_{t} \log \left(Q_{t}\right)-\left(1-\bar{Q}_{t}\right) \log \left(1-Q_{t}\right)\right) \\
     \mathcal{L}_{\mathrm{dtd}}&=\mathcal{L}_{\mathrm{dtd-nearend}} + \mathcal{L}_{\mathrm{dtd-farend}} \\
     \end{split}
% \right.
\end{equation}
}%
where $t$ and $f$ denote frame and frequency respectively, and $\bar{P}_{t},\bar{Q}_{t} \in\{0,1\}$ are the near-end and far-end speech activity label respectively, which is based on a short-term energy threshold. ${P}_{t}$ and ${Q}_{t}$ are the estimated state of near-end and far-end speech, respectively.

The final loss is
{
\setlength\abovedisplayskip{2.0pt}
\setlength\belowdisplayskip{2.0pt}
\begin{equation}
\begin{split}
\footnotesize
    % \vspace{-0.1cm}
% \left \{
     \setlength{\arraycolsep}{0.3pt}
     \mathcal{L}_{\mathrm{final}}&=\mathcal{L}_{\mathrm{echo-aware}} + 0.2\mathcal{L}_{\mathrm{mask}} + 0.1\mathcal{L}_{\mathrm{dtd}} \\
&+ 0.05\mathcal{L}_{\mathrm{echo}} + \mathcal{L}_{\mathrm{asym}}
% \right.
% \vspace{-0.1cm}
\end{split}
\end{equation}
}%
where the definition of $\mathcal{L}_{\mathrm{mask}}$ is the same as that in~\cite{zhang2022multi} and $\mathcal{L}_{\mathrm{asym}}$ is consistent with that in~\cite{braun2022task}.

\vspace{-10pt}
\section{Experiments}
\vspace{-6pt}
\label{sec:typestyle}
\subsection{Data and experiment setup}
\vspace{-3pt}
In our experiment, the clean speech dataset is from the ICASSP 2022 DNS-challenge~\cite{dubey2022icassp}. The noise data originates from AudioSet, Freesound, and Demand databases. We also use the far-end single-talk clips officially provided by the AEC challenge as echo signals for training and validation. For RIR, we generate 100,000 pairs using the image method for the training set and 5,000 for the validation set, respectively. The signal-to-noise ratio (SNR) and signal-to-echo ratio (SER) are set to [0, 20]dB and [-10, 15]dB, respectively, for generating near-end microphone signals. SNR is set to [15, 45]dB for far-end signal. We create a small dataset (300 hours) for the ablation study and a large dataset (1400 hours) for final submission model training. In the ablation study, a simulated test set of 1500 clips is used for model evaluation, 400 clips for far-end single talk, 400 clips for near-end single talk, and the rest are double talk.

For the proposed model, the window length and frameshift are 20ms and 10ms respectively. As the computational complexity of the full-band model exceeds the requirements of the challenge, the sub-band approach is taken and the full-band signals are processed into 4 sub-band signals using PQMF. All neural models are trained with the Adam optimizer for 60 epochs with an initial learning rate of 1e-4, and the learning rate is halved if there is no loss decrease on the development set for 2 epochs.

For the output of each layer of encoder/decoder, the number of channels is 80. The kernel size and stride of GConv and TrGConv are set as (2, 3), (1, 2) in the time and frequency axis, respectively. In the U-TFCM module, 4 convolutional layers are adopted with a dilation rate of $\{1, 2, 4, 8\}$ in both the time axis and frequency axis. The stride and kernel size of DConv in U-TFCM are set as (1, 1), (3, 3). The TFCM module is the same as that in ~\cite{ju2023tea}. The number of parameters of our submitted model is 3.83M. The RTF of the system is 0.1983 running on Intel(R) Xeon(R) Silver 4214R CPU @ 2.40GHz while the neural post-filter is implemented by ONNX for speed-up.
\vspace{-10pt}
\subsection{Results and analysis}
\label{sec:majhead}
\vspace{-3pt}
We first perform an ablation study on a small data set (300h). Based on the results in Table~\ref{tab:test}, it is evident that incorporating the DSVAD
module results in improvements in all aspects, especially the performance of far-end single-talk suppression. Additionally, integrating the U-TFCM and TFCM modules into the encoder and decoder exhibits extra performance gain, resulting in a further 0.15 WB-PESQ gain for the DT scenario. Moreover, auxiliary training with the echo decoder significantly enhances echo suppression. Finally, we train the whole model using the large dataset (1400h) and process the blind test clips as our submission.

% From Table~\ref{tab:test}, we first see that adding the DSVAD module can slightly improve the model in all aspects. Second, the U-TFCM and TFCM modules to the encoder and decoder parts demonstrated their excellent modeling ability, achieving 0.15 improvement on DT scenario. Finally, according to our experimental results, the echo decoder auxiliary training helps echo suppression, we process the blind test clips using this model.

From the blind test set results in Table~\ref{tab:rtf}, we
can observe that our S-TFGCRN model significantly outperforms the baseline, ranking 4th in the non-personalized track according to the official results.

    \begin{table}[!h]
    \vspace{-15pt}
    \centering
     % \vspace{-0.1cm}
    \footnotesize
    \setlength\tabcolsep{4pt}% 调整列间距
    \caption{Echo suppression performance in the simulated test set. DT: double talk, ST: single-talk, NE: near-end, FE: far-end.}
%     WB-PESQ used for DT and ST-NE
% scenarios and ERLE used for ST-FE scenario in both simulated test set (ST-FE).
    \vspace{2pt}
     \resizebox{\linewidth}{!}{
    \begin{tabular}{lccccc}
    
    \toprule 
    Model & Para.(M) & \makecell[c]{DT\\(WB-PESQ)} & \makecell[c]{ST-NE\\(WB-PESQ)} & \makecell[c]{ST-FE\\(ERLE)} & Data\\ 
    % \midrule 
    %  & & WB-PESQ & WB-PESQ & ERLE\\
    \midrule 
    Noisy & & 2.03 & 2.85 & 0 & \multirow{6}{*}{300h}\\
    Sub-band GCRN & 2.34 & 2.84 & 3.29 & 56.58\\
    ~ + DSVAD & 2.57 & 2.87 & 3.31 & 62.82\\
    % ~ ~ + Asym loss & 2.57 & 2.81 & 3.25 & 57.36\\
    ~ ~ + U-TFCM, TFCM & 3.14 & 3.02 & 3.40 & 63.83\\
    ~ ~ ~ + Echo Decoder & 3.83 & 3.06 & 3.44 & 66.27\\
    \midrule
    S-TFGCRN (submitted) & 3.83 & \textbf{3.16} & \textbf{3.48} & \textbf{67.43} & 1400h\\      
    \bottomrule
    \label{tab:test}
    \end{tabular}
    }
    \vspace{-20pt}
    \end{table}

    \begin{table}[!h]
    \vspace{-15pt}
    \centering
     \vspace{5pt}
    \footnotesize
    \setlength\tabcolsep{5pt}% 调整列间距
    \caption{Model performance on the blind test set.} 
    \vspace{5pt}
    \resizebox{\linewidth}{!}{
    \begin{tabular}{lccc}
    \toprule 
    Model & Overall MOS & WAcc & Final Score\\ 
    \midrule 
    Baseline & 4.013 & 0.649 & 0.736 \\ 
    S-TFGCRN (submitted) & \textbf{4.320} & \textbf{0.790} & \textbf{0.823} \\ 
    \bottomrule
    \label{tab:rtf}
    \end{tabular}
    }
    \vspace{-18pt}
    \end{table}

%\vspace{-12pt}
%\section{Conclusion}
%\vspace{-8pt}
%\label{sec:page}
%In this paper, we introduce our entry to the ICASSP 2023 AEC Challenge for non-personalized echo cancellation. Specifically, sub-band processing is adopted to reduce the computational complexity of the model. Our model's ability for echo cancellation is enhanced with the help of the encoder, decoder, TFCM, and U-TFCM. Multi-task learning with DSVAD and echo estimation module can achieve good echo cancellation and prevent excessive suppression of near-end speech.
%Our proposed method manages to rank 4th place in the challenge with good subjective quality (MOS) and speech recognition accuracy (WAcc).
%\vfill\pagebreak

% \section{REFERENCES}
% \label{sec:refs}

% List and number all bibliographical references at the end of the
% paper. The references can be numbered in alphabetic order or in
% order of appearance in the document. When referring to them in
% the text, type the corresponding reference number in square
% brackets as shown at the end of this sentence \cite{C2}. An
% additional final page (the fifth page, in most cases) is
% allowed, but must contain only references to the prior
% literature.

\vspace{-10pt}
% \footnotesize
\bibliographystyle{IEEEbib}
\bibliography{strings,refs}

\end{document}